\documentstyle[12pt,epsf]{article}
\begin{document}

\par
\topmargin=-1cm      

{ \small
\noindent{U.Md. PP\# 98-024}\hfill{DOE/ER/40762-131}

}

\vspace{40.0pt}

\begin{center}
{\large {\bf The Potential of Effective Field Theory in $NN$ Scattering}}\\

\vspace{12pt}
{\bf S.R.~Beane, T.D.~Cohen and D.R.~Phillips\footnote{Email: 
sbeane,cohen,phillips@quark.umd.edu}}\\
\vspace{12pt}
Department of Physics, University of Maryland, College Park, MD, 
20742-4111

\end{center}
\vspace{6pt}

\begin{abstract}
{\normalsize We study an effective field theory of interacting
nucleons at distances much greater than the pion's Compton
wavelength. In this regime the $NN$ potential is conjectured to be the
sum of a delta function and its derivatives. The question we address
is whether this sum can be consistently truncated at a given order in
the derivative expansion, and systematically improved by going to
higher orders.  Regularizing the Lippmann-Schwinger equation using a
cutoff we find that the cutoff can be taken to infinity only if the
effective range is negative. A positive effective range---which occurs
in nature---requires that the cutoff be kept finite and below the
scale of the physics which has been integrated out, i.e.~$O(m_\pi)$.
Comparison of cutoff schemes and dimensional regularization reveals
that the physical scattering amplitude is sensitive to the choice of
regulator. Moreover, we show that the presence of some regulator
scale, a feature absent in dimensional regularization, is essential if
the effective field theory of $NN$ scattering is to be useful. We
also show that one can define a procedure where finite cutoff
dependence in the scattering amplitude is removed order by order in
the effective potential. However, the characteristic momentum in the
problem is given by the cutoff, and not by the external momentum. It
follows that in the presence of a finite cutoff there is no small
parameter in the effective potential, and consequently no systematic
truncation of the derivative expansion can be made. We conclude that
there is no effective field theory of $NN$ scattering with nucleons
alone.}

\end{abstract}

\newpage

\section{Introduction}

There exist many nucleon-nucleon potentials which reproduce phase
shifts and nuclear properties with remarkable
accuracy~\cite{Na78,La80,Ma87,Ma89,St94,Wi95,Ma96,St97}. Three
fundamental features are shared by these potential models: (i) pions
dominate at long distances, (ii) there is some source of
intermediate-range attraction, and (iii) there is some source of
short-distance repulsion. However, in general, distinct physical
mechanisms in these models account for the same feature of the nuclear
force.  Agreement with experiment is maintained in spite of these
differences because of the large number of fit parameters.  It would
be a considerable advance if a systematic approach to the
nucleon-nucleon interaction could be developed based solely on
symmetries and general physical principles.

Systematic approaches to the scattering of strongly interacting
particles, such as chiral perturbation theory, are based on the ideas
of effective field theory (EFT). Effective field theory says that for
probes of a system at momentum $k \ll M$, details of the dynamics at
scale $M$ are unimportant. What is important at low energies is the
physics that can be captured in operators of increasing dimensionality
which take the form of a power-series in the quantity
$k/M$~\cite{Ka95,Ma95}. It is important to realize that even if EFT
ideas can be applied to the $NN$ system they will probably not lead to
any startlingly new predictions for $NN$ scattering. Indeed, it is
entirely possible that the resulting fits to phase shifts will not be
as good as those produced by conventional $NN$ potentials with the
same number of parameters.  However, the real motivation for
constructing an EFT is to relate one process to another.  For
instance, one would like to relate $NN$ scattering systematically to
scattering processes with more nucleons, such as $NNN$ and $NNNN$
scattering, and to say something predictive about processes involving
pionic and photonic probes of few-nucleon systems.

One reason to hope that this can be achieved in nuclear physics is
provided by the pattern of chiral symmetry breaking in QCD. The fact
that chiral symmetry is spontaneously broken implies that the pion is
light and interacts weakly at low energies. Of course the lightness of
the pion in itself guarantees that it should play a fundamental role
in nuclear physics. The weakness of pion interactions at low energies
allows pion interactions with nucleons to be systematized using chiral
perturbation theory. This procedure has proved remarkably successful
in describing the interactions of pions with a single
nucleon~\cite{Br95}. These ideas have been extended to processes
involving more than one nucleon, leading to arguments concerning the
relative importance of various pion-exchange mechanisms in the nuclear
force~\cite{We90,We91,Fr96}. However, these EFT arguments assume the
existence of a systematic power-counting scheme for multi-nucleon 
processes. In our view there is no convincing
argument in the literature that such a power counting exists.  The
purpose of this paper is to investigate this issue in the simplest
possible context.

Naive transposition of EFT ideas to nuclear physics immediately
suggests a puzzle.  In nuclear physics there are bound states whose
energy is unnaturally small on the scale of hadronic physics. In order
to generate such bound states within a ``natural'' theory it is clear
that one must sum a series to all orders.  Therefore, Weinberg
proposed~\cite{We90,We91} implementing the EFT program in nuclear
physics by applying the power-counting arguments of chiral
perturbation theory to an $n$-nucleon effective potential rather than
directly to the S-matrix. Only $n$-nucleon irreducible graphs should
be included in the $n$-nucleon effective potential. The potential
obtained in this way is then to be inserted into a Lippmann-Schwinger
or Schr\"odinger equation and iterated to all orders. There will of
course be unknown coefficients in the effective potential, but these
can be fit to experimental data as in ordinary chiral perturbation
theory~\cite{Or96,Ka96,Sc97,Le97} .

Thus, an EFT treatment of the $NN$ interaction differs in a
fundamental way from conventional EFT applications like $\pi \pi $
scattering in chiral perturbation theory. In both cases operators are
ordered in an effective Lagrangian in the same way. However, in $\pi
\pi $ scattering the operator expansion in the effective Lagrangian
maps to a power series in $k/M$ in the scattering amplitude. It is
straightforward to see that EFT treatments where there is a direct
mapping from the Lagrangian to the S-matrix are
systematic~\cite{We79}. On the other hand, when the mapping is from
the Lagrangian to an effective potential which is subsequently
iterated to all orders, many issues arise which lead one to question
the existence of a systematic power counting in the potential.

In order to raise some of these issues, consider $NN$ scattering in
the ${}^1S_0$ channel at momentum scales $k \ll m_\pi$.  The EFT at
these scales involves only nucleons since the pion is heavy and may
therefore be ``integrated out''.  The effective Lagrangian thus
consists of contact operators of increasing dimensionality constrained
by spin and isospin. Throughout the paper this is the EFT which we
consider. We do not intend that this EFT should provide a quantitative
description of the $NN$ phase shifts. Instead, we study it because 
scattering amplitudes can be calculated analytically. It therefore
allows us to elucidate issues of principle in EFT for $NN$ scattering.

\begin{figure}[h,t,b]
   \vspace{0.5cm} 
   \epsfysize=2cm 
   \centerline{\epsffile{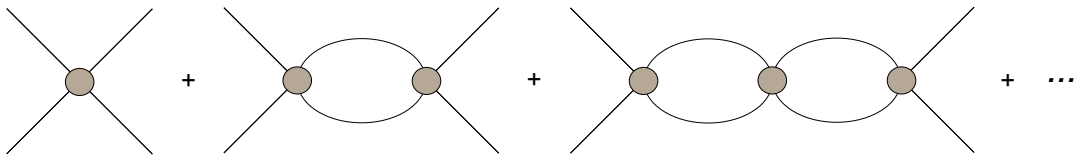}}
   \centerline{\parbox{11cm} {\caption{\label{fig1} The diagrammatic
   solution of the LS equation with the effective potential
   represented by the shaded blob.  }}}
\end{figure}

One might naively expect to be able to calculate the $NN$ scattering
amplitude directly from the effective Lagrangian as a power series in
$k/ {m_\pi}$, where $m_\pi$ is the heavy scale in this problem. It is
instructive to show in some detail why this fails and one is led to
consider an expansion in the potential. Consider an expansion of the
amplitude in the $S$-wave channels:

\begin{equation}
T^{\rm on}(k,k)=c_0 + c_2 \left(\frac{k}{m_\pi}\right)^2 + c_4
\left(\frac{k}{m_\pi}\right)^4 + \ldots,
\label{eq:Texp}
\end{equation}
where $k^2=ME$, and, with the prevailing prejudice of EFT, we
anticipate that the dimensionless coefficients $c_0$, $c_2$, etc. will
be natural; i.e. of order unity. We know that in the ${}^1S_0$ and
${}^3S_1-{}^3D_1$ channels there are, respectively, a quasi-bound
state and a bound state at low energies. The power series expansion
(\ref{eq:Texp}) with natural coefficients can only be correct if these
bound states are at energies $k^2 \sim m_\pi^2$. However, in these
channels the bound states occur at unnaturally low energies, i.e. at
energies $k^2 \ll m_\pi^2$. Therefore, the coefficients in the
expansion must be unnatural---they are fixed by the pole positions of
the low-lying bound states rather than by the scale of the physics
that has been integrated out.  This limits the usefulness of an
expansion in the amplitude to an extremely restricted domain of
validity.  On the other hand, if one makes the following expansion of
the potential in $S$-wave channels:

\begin{equation}
V(p',p)=c_0 + c_2 \frac{p^2 + p'^2}{m_\pi^2}+ c_4
\frac{p^4 + p'^4}{m_\pi^4} + c_4' \frac{p^2 p'^2}{m_\pi^4}+\dots,
\label{eq:Vexp}
\end{equation}
and iterates it via the Lippmann-Schwinger equation (see Fig.~\ref{fig1})

\begin{equation}
T(p',p;E)=V(p',p) + M\int \frac{d^3q}{(2 \pi)^3} \, V(p',q) 
\frac{1}{EM- {q^2}+i\epsilon} T(q,p;E),
\label{eq:LSE}
\end{equation}
one may hope to generate (quasi-)bound states at the appropriate
energies while maintaining natural coefficients in the potential. At
face value this procedure appears promising.  The expansion
(\ref{eq:Vexp}) may be truncated at some finite order in the
quantities $p/m_\pi$ and $p'/m_\pi$ and, provided $p,p' \ll m_\pi$ the
neglected terms will be small.  However, in making the expansion in
the potential of Eq.~(\ref{eq:Vexp}) and then iterating, a number of
issues arise which are absent in standard EFT treatments, and which
render this procedure suspect. In what follows we will identify and
clarify some of these issues.

First, the physical scattering amplitude that is generated when the
potential (\ref{eq:Vexp}) is iterated is exactly unitary and therefore
necessarily contains arbitrarily high powers in energy or momenta.
This occurs regardless of the order to which one is working in the
momentum expansion of the potential $V$.  This suggests that as long
as one is interested in physics near the bound state pole---where
exact unitarity is important---the scattering amplitude may be
sensitive to physics at arbitrarily short-distance scales. While this
observation does not necessarily invalidate the EFT approach, it does
mean that one should not simply translate intuition gained about how
short-distance physics decouples in perturbative EFT calculations
(e.g. in chiral perturbation theory) to the non-perturbative problem
at hand.  Thus, the first question we face is whether a systematic
expansion in the potential translates to a systematic determination of
the scattering amplitude. A necessary condition for this to occur is
that the integrals in the LS equation, which formally extend up to
infinity, be dominated by momenta of order $p \sim k \ll m_\pi$, where
$k$ is the on-shell momentum, since otherwise there is no small
parameter in the expansion (\ref{eq:Vexp}) for $V$.

Second, non-perturbative regularization and renormalization is
required when iterating to all orders using the LS equation.  This is
an issue because of the presence of ultraviolet divergences, which
generally can arise in two ways.  When explicit pions are included in
the EFT, the potential itself may contain loop graphs which require
regularization and renormalization. These divergences are easily
handled using standard perturbative methods.  A second type of
ultraviolet divergence arises because in solving the LS equation one
integrates the potential over all momenta. Given the hard asymptotic
behavior inherent to the momentum expansion this necessarily
introduces new divergences. The divergences which arise from iterating
the potential become more severe as one goes to higher order in the
EFT expansion.  These divergences apparently violate the assumption
that $p\ll m_\pi$, so the existence of a procedure to regularize these
divergences and renormalize in such a way that the momentum scales
probed inside loops are ultimately well below $m_\pi$ provides a
non-trivial condition on the existence of an EFT.

In ordinary perturbative EFT, all regularization schemes lead to the
same renormalized amplitude. This insensitivity to the short-distance
physics implied by the regulator makes perturbative EFT methods like
chiral perturbation theory possible. Similarly, there is no hope of
defining a sensible EFT for $NN$ scattering unless there is some
degree of regulator independence.  An interesting feature that arises
when regularizing and renormalizing the Lippmann-Schwinger equation is
that not all regulators lead to the same physical results.  For
instance, we will show that dimensional regularization (DR) and cutoff
schemes lead to different physical scattering amplitudes. This result
should give practitioners of EFT pause, since it suggests a
sensitivity to short-distance physics which violates the basic tenets
of EFT. We further demonstrate that in order to generate low-energy
(quasi-)bound states in the $NN$ system within this EFT, one must use
a scale-dependent regulator.

Cutoff schemes provide the most physical means of regularizing the
effective theory. We therefore focus on whether it is possible to
implement the regularization and renormalization program in a
consistent fashion using cutoff schemes. Within this restricted class
of regulators physical results are insensitive to the specific choice
of regulator.  However, we find that taking the cutoff to infinity
requires that the effective range parameter in the scattering
amplitude be negative. In physical processes of interest the effective
range parameter is positive. The impossibility of maintaining a
positive effective range when the cutoff is removed follows from an old
theorem of Wigner, which depends only on general physical
principles~\cite{Wi55,PC97}. Therefore, either the cutoff must be kept
finite or else all orders in the effective potential must be retained.
Using an extension of Wigner's theorem to the EFT where pions are
explicitly included, we further argue that the inclusion of pions as
explicit degrees of freedom in the EFT does not resolve these
difficulties.

Having found that an EFT with only nucleons as explicit degrees of
freedom can work only if there is a finite cutoff or regulator scale,
we investigate the possibility of a cutoff effective field theory.  In
cutoff EFT, cutoff dependence in physical observables is removed
systematically by adding higher-dimensional operators to the effective
action. However, this procedure makes sense only if there is a small
parameter which allows one to conclude that higher-order operators in
the action are negligible. We show that there is no such small
parameter in $NN$ scattering when there are low-energy (quasi-)bound states.

We stress that we {\it are not} questioning the existence of an
effective field theory of $NN$ scattering. Since the principles
underlying EFT are causality and locality, an EFT of $NN$ scattering
must exist unless a sacred principle is violated.  The question we
address is: ``What are the relevant low-energy degrees of freedom
which must be included explicitly in the EFT?'' The problem of $NN$
scattering is subtle in this respect because generally there are
singularities at unnaturally low energies (e.g., the deuteron) in
scattering amplitudes. These singularities are not present as fields in
the effective action.  What our analysis here shows is that if
low-lying bound states are present then there is no systematic EFT for
$NN$ scattering in which there are only nucleon fields in the
effective Lagrangian. This would suggest the necessity of introducing
the physics of (quasi-)bound states in terms of explicit degrees of freedom
in the EFT description. A step in this direction has been taken in
Ref.~\cite{Ka97}, but in this paper we will not discuss such ideas
further. Therefore, we use the term EFT to mean an effective field
theory in which low-energy bound states are not included as fields in
the effective action, i.e. an EFT in which only nucleons (and pions)
are explicit degrees of freedom.

In Sec.~2 we discuss the ${}^1S_0$ channel in $NN$ scattering and
introduce a simple potential model that reproduces data remarkably
well. This model demonstrates how an unnaturally small scale can
emerge from a theory with only large scales.  In Sec.~3 we review the
effective field theory power-counting argument for $NN$ scattering,
originally proposed by Weinberg. In particular, we examine arguments
for power counting in the potential rather than in the amplitude in
the $NN$ problem.  In Sec.~4 we perform a ``leading-order'' EFT
calculation for $NN$ scattering in the ${}^1S_0$ channel and show that
all regularization schemes give the same scattering amplitude.  In
Sec.~5 we perform a ``second-order'' EFT calculation using cutoff
regularization.  We show that if a cutoff is introduced such that the
regulated theory respects all physical principles, then the cutoff
cannot be taken to infinity.  We discuss how this result relates to an
old theorem of Wigner. In Sec.~6 we compare dimensional regularization
(DR) and cutoff regularization schemes. We show that DR and cutoff
regularization give different physical scattering amplitudes in the
second-order EFT calculation, indicating that there is a sensitivity
to the choice of regulator, and therefore to short-distance physics.
We find that the presence of a regulator scale is necessary to
reproduce unnaturally low-lying (quasi-) bound states. In Sec.~7 we
consider the possibility of an effective field theory with a finite
cutoff. We find that although a procedure whereby cutoff dependence in
the amplitude is removed by adding higher orders in the effective
potential can be defined, the small parameter which is necessary to
conclude that this is systematic is simply not present.  We summarize
and conclude in Sec.~8.

\section{Phenomenology of the ${}^1 S_0$ $NN$ Interaction}
\label{sec-sm}

\begin{figure}[h,t,b]
   \vspace{0.5cm} 
   \epsfysize=8 cm
   \centerline{\epsffile{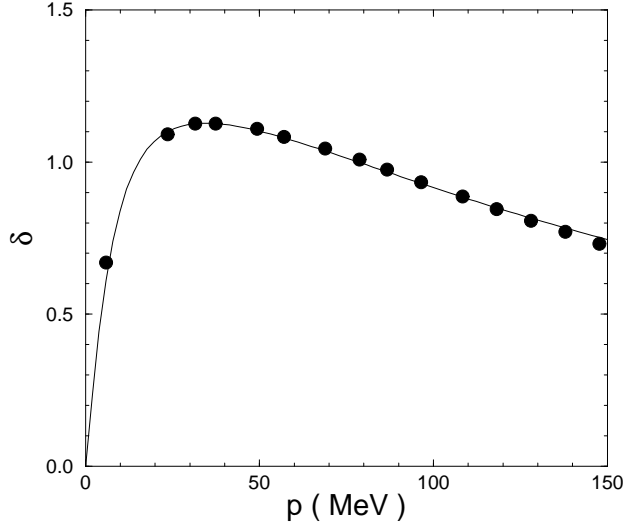}}
   \centerline{\parbox{11cm}
   {\caption{\label{fig2} The effective
range expansion with the extracted phase-shift data up 
to center-of-mass momenta of order $m_\pi$. Phase shift data
are taken from Ref.~\protect\cite{St93}.}}}
\end{figure}

The ${}^1 S_0$ phase shift in $NN$ scattering is remarkably well
reproduced (see Fig.~\ref{fig2}) up to center of mass momenta of order
$m_\pi$ by the first two terms in the effective-range expansion:

\begin{equation}
\frac{1}{T^{\rm on}(k)}=-\frac{M}{4 \pi}\left[ -\frac{1}{a} +
\frac{1}{2} r_e {k^2} - i k \right],
\label{eq:reamp}
\end{equation}
where $a$ is the scattering length, $r_e$ is the effective range, and
$k$ is the on-shell momentum, $k=\sqrt{ME}$. Experimentally these
parameters are determined to be

\begin{equation}
a=-23.714\pm 0.013\, {\rm fm}\qquad {r_e}=2.73\pm 0.03\, {\rm fm}.
\label{eq:cexpo2}
\end{equation}
The extremely large (negative) value of the scattering length implies
that there is a virtual bound state in this channel very near zero energy.
While the value of $r_e$ is consistent with what one might expect
for a ``natural'' theory where pions dominate the low-energy physics
($r_e \sim 1/m_\pi$), the value of $a$ is far from being natural 
($a \gg 1/m_\pi$).

In order to understand the emergence of an unnaturally large
scattering length it is instructive to consider a simple separable
potential model whose scattering amplitude exactly reproduces the
effective-range expansion, Eq.~(\ref{eq:reamp}), with no terms of higher
order in momenta. Consider the potential

\begin{equation}
V(p',p)=-\frac{4\pi g}{\Lambda M}(1+\frac{p'^2}{\Lambda^2})^{-1/2}
(1+\frac{p^2}{\Lambda^2})^{-1/2},
\end{equation}
where $g$ is a dimensionless coupling constant and $\Lambda$ is a
scale that characterizes the range of the potential. Clearly this
model has nothing do with effective field theory.  It is a simple
model for the ${}^1S_0$ channel which captures some of the salient
features of the scattering.

Inserting this potential into the LS equation (\ref{eq:LSE}) gives
the on-shell amplitude:

\begin{eqnarray}
\frac{1}{T^{\rm on}(k)}&=&
{V(k,k)^{-1}}\left[1-M\int \frac{d^3q}{(2 \pi)^{3}} \, \frac{V(q,q)}
{{k^2}- {q^2}+i\epsilon}\right] \nonumber \\
&=&-\frac{M}{4\pi}\left[ \frac{\Lambda}{g}(1+\frac{k^2}{\Lambda^2})
-(\Lambda +ik)\right].
\label{eq:beanemodel}
\end{eqnarray}
This is simply the effective-range expansion to order
$k^2$.  One can easily read off the scattering length and effective
range, by comparing Eqs.~(\ref{eq:beanemodel}) and (\ref{eq:reamp}):

\begin{equation}
a= \frac{1}{\Lambda} \frac{g}{g-1} \qquad {r_e}=\frac{2}{g\Lambda}.
\label{ex}
\end{equation}
The choices $\Lambda =152$ MeV and $g=0.95$ then reproduce the values
(\ref{eq:cexpo2}). 

Note that the effective range is directly proportional to the range,
$1/\Lambda$, of the potential. The expression for $a$ makes clear that
an unnaturally large scattering length occurs if $g$ is
``accidentally'' close to one~\footnote{See also~\cite{Ka97} for an
illustrative example.}. In this model the large scattering length is
the result of a cancellation between two distinct contributions to the
scattering amplitude. The first of these contributions comes from the
``strength'' of the potential (the blob in Fig.~\ref{fig1}) since it
originates in the $1/V(k,k)$ term in Eq.~(\ref{eq:beanemodel}).  The
second term comes from virtual excitations in the loops in
Fig.~\ref{fig1}. This contribution is sensitive to the range of the
potential, i.e. $1/\Lambda$.  Note that both of these contributions to
$1/a$ are natural, i.e. given by $-\Lambda/g$ and $\Lambda$
respectively.  Nevertheless, if $g \approx 1$ there is a cancellation
between the ``strength'' and the ``range'', which results in an
unnaturally large scattering length. We will see that this qualitative
feature persists in the EFT treatment.

\section{An Effective Field Theory for $NN$ Scattering}
\label{sec-eft}

In its standard form chiral perturbation theory is not useful in $NN$
scattering because of the presence of low-lying bound states which
imply a breakdown of perturbation theory~\cite{We90,We91}.  Weinberg
argued that this breakdown manifests itself through the presence of
infrared divergences in reducible loop graphs in the limit of static
heavy nucleons. This argument is not entirely correct---it {\it is}
possible to have a ``perturbative'' effective field theory for the
interaction of heavy particles, provided that there are no low-lying
bound or quasi-bound states. (See, for instance Ref.~\cite{LM97}.)
However, since there {\it are} low-lying bound states in nuclear
systems, Weinberg's main point remains true: if power-counting
arguments are relevant in nuclear physics, they must be applied to an
effective potential, rather than to the scattering amplitude.  In
general, the effective $NN$ potential is the sum of all
two-particle-irreducible (2PI) $NN \rightarrow NN$ diagrams.  When
there are pions and nucleons in the EFT, the effective potential
contains pion-exchange diagrams, short-range contact interactions and
a mixture of the two.  However, in this paper we are interested in
energy scales where $m_\pi$ is large compared to the initial and final
momenta $p$ and $p'$.  In this case EFT arguments imply that the pion
can be integrated out and replaced by contact interactions. 

The starting point of any EFT calculation is an effective Lagrangian,
where operators are ordered according to the number of derivatives.
The most general effective Lagrangian consistent with spin and
isospin, including only operators relevant to ${}^1S_0$ scattering is

\begin{equation}
{\cal L}=N^\dagger i \partial_t N - N^\dagger \frac{\nabla^2}{2 M} N
- \frac{1}{2} C (N^\dagger N)^2\\ 
-\frac{1}{2} C_2 (N^\dagger \nabla^2 N) (N^\dagger N) + h.c. + \ldots.
\label{eq:lag}
\end{equation}
There are only tree graphs in the $NN$ potential obtained from this
Lagrangian. Hence there is a one-to-one correspondence between the
operators in the effective Lagrangian and the terms in the
potential. When iterated using the Lippmann-Schwinger equation, this
potential gives the full $NN$ amplitude of the theory. This procedure
is entirely equivalent to solving the quantum partition function with
an effective action to a given order in the derivative expansion, and
thereby generating the four-point correlation function. There is
therefore, at least for this EFT, actually no need to formulate the
discussion in terms of potentials. Nevertheless, here we use the
familiar language of non-relativistic quantum mechanics.

The effective potential to order $\nu$ in the derivative
expansion can be expressed in the form~\cite{We90,We91,Fr96,Or96}

\begin{equation}
V^{(\nu)}(p',p)=\frac{1}{\Lambda^2} \sum_{{\bar\nu}=0}^{\nu}
\left[\frac{(p,p')}{\Lambda}\right]^{\bar\nu} c_{\bar\nu}
\label{eq:chiralexp}
\end{equation}
where the sum here is over all possible terms extracted from
(\ref{eq:lag}) and $\Lambda$ is the scale of the physics integrated
out, taken to be $m_\pi$ in the present context. Equation
(\ref{eq:chiralexp}) is intended to be symbolic; $(p,p')$ indicates
that either of these quantities may appear in the expansion, in any
combination consistent with symmetry, with only their total power
constrained. For instance, at $\bar\nu =4$ we have the structures $p^4
+p'^4$ and ${p^2}{p'^2}$ with coefficients $c_4$ and $c_4'$,
respectively. It is assumed that the coefficients $c_{\bar\nu}$ are
natural; that is, of order unity.  The fundamental assumption
underlying effective field theory for the $NN$ interaction is that
this expansion in the potential, or equivalently that in the
Lagrangian, may be sensibly truncated at some finite order, $\nu$. It
is the purpose of this paper to test this assumption.

The physical scattering amplitude is obtained by iterating
the potential (\ref{eq:chiralexp}) using the Lippmann-Schwinger equation

\begin{equation}
T(p',p;E)=V(p',p) + M\int \frac{d^3q}{(2 \pi)^3} \, V(p',q) 
\frac{1}{EM- {q^2}+i\epsilon} T(q,p;E),
\label{eq:LSE2}
\end{equation}
which generates the $T$-matrix. This procedure is illustrated in
Fig.~\ref{fig1}. By assumption, truncating the expansion
(\ref{eq:chiralexp}) and retaining only its first few terms will be
valid only for nucleon momenta well below $\Lambda$.  It is clear that
a method of regularizing the otherwise-divergent integrals which occur
when potentials such as (\ref{eq:chiralexp}) are inserted into the LS
equation must be specified.

\section{``Leading order'' EFT calculation ($\nu=0$)}
\label{sec-nu0}

We first investigate these questions by considering the ${}^1S_0$ $NN$
amplitude generated by the ``zeroth-order'' EFT potential.  This
potential is

\begin{equation}
V^{(0)}(p',p)=C.
\end{equation}
The solution of the LS equation with
this potential is easily found to be

\begin{equation}
\frac{1}{T^{\rm on}(k)}=\frac{1}{C} - I(k),
\end{equation}
where $k=\sqrt{ME}$ is the on-shell momentum, and 

\begin{equation}
I(k) \equiv M\int \frac{d^3q}{(2 \pi)^{3}} \, \frac{1}
{{k^2}- {q^2}+i\epsilon}.\label{eq:IEdef}
\end{equation}
This integral is linearly divergent. Regularizing with a sharp
cutoff, $\beta$, gives

\begin{equation}
\mbox{Re} \left(\frac{1}{T^{\rm on}(k)}\right)=\frac{1}{C_{\rm cutoff}} + 
\frac{M \beta}{2\pi^2},
\label{eq:T0cutoff}
\end{equation}
where we have neglected terms that vanish as powers of $1/\beta$.
We choose the renormalization condition

\begin{equation}
\mbox{Re} \left(\frac{1}{T^{\rm on}(0)}\right) \equiv \frac{1}{C_{\rm ren}}=
\frac{M}{4\pi a}
\label{eq:renormcond}
\end{equation}
where $C_{\rm ren}$ is the ``renormalized potential'' and $a$ is the
physical scattering length. Note that just as in the example of
Sec.~\ref{sec-sm} there is a cancellation in Eq.~(\ref{eq:T0cutoff})
between the contributions to $1/a$ from the strength of the potential,
and the integration over loop momenta.

In the limit $\beta\rightarrow\infty$ we then find
the physical renormalized scattering amplitude,

\begin{equation}
\frac{1}{T^{\rm on}(k)}=
-\frac{M}{4 \pi}\left[ -\frac{1}{a} - i k \right],
\label{eq:T0}
\end{equation}
derived by Weinberg~\cite{We91} and Kaplan {\it et al.}~\cite{Ka96}.
Note that this result is completely insensitive to choice of
regularization scheme. After renormalization the only piece of the
integration over internal loop momenta which survives is the imaginary
part, which gives the unitarity branch cut and depends only on the
on-shell momentum, $k$.  The real piece of $I(k)$, which comes from
integration over virtual nucleon momenta, is absorbed into the
constant $1/C_{\rm{cutoff}}$.  This is exactly how one renormalizes in
standard EFT, where power-law divergent pieces of loop integrals are
absorbed by bare coefficients in the amplitude.  One efficient way to
ensure that the power-law divergent parts of loop graphs never enter
the calculation is to use dimensional regularization. However, note
that if DR is used, as in Ref.~\cite{Ka96}, then the cancellation seen in
Eq.~(\ref{eq:T0cutoff}) is destroyed, since the linear divergence is zero
in DR. Although of no physical consequence here, this effect will
prove important at second order in the EFT.

The result (\ref{eq:T0}) appears promising: we have reproduced the
scattering length term of the effective-range expansion. Moreover, we
see that at ``leading order'' in the effective field theory
calculation the scattering amplitude is regularization-scheme
independent; the real part of the integration over loop momenta does
not have any effect on the final physical result. In order to test the
robustness of this result we must calculate corrections. At the next
order the potential has momentum dependence and therefore is more
sensitive to short-distance physics.

\section{``Next-to-leading Order'' EFT calculation($\nu$=2)}
\label{sec-nu2}

{}From the effective Lagrangian (\ref{eq:lag}) we extract the following
``second-order'' EFT potential:

\begin{equation}
V^{(2)}(p',p)=C + C_2(p^2 + p'^2).
\label{eq:V2}
\end{equation}
An easy way to solve the LS equation with this potential is to observe
that $V^{(2)}$ may be written as a two-term separable potential. The
on-shell T-matrix is easily found to be (see Ref.~\cite{Ph97} for details):

\begin{equation}
\frac{1}{T^{\rm on}(k)}=\frac{(C_2 I_3 -1)^2}{C + C_2^2 I_5 + {k^2} C_2 (2 -
C_2 I_3)} - I(k),
\label{eq:Tonexp}
\end{equation}
where

\begin{equation}
I_n \equiv -M \int \frac{d^3q}{(2 \pi)^3} q^{n-3}.
\label{In}\end{equation}
The integrals $I_3$, $I_5$ and $\mbox{Re}(I(k))={I_1}$ are all divergent,
and so this scattering amplitude must be regularized and renormalized.
It is instructive to carry the renormalization a certain distance
without specifying a regularization scheme.  We choose as renormalized
parameters the experimental values of the scattering length, $a$, and the
effective range, $r_e$. In other words, we fix $C$ and $C_2$ by
demanding that

\begin{equation}
\frac{1}{T^{\rm on}(k)}=-\frac{M}{4 \pi}\left[ -\frac{1}{a} +
\frac{1}{2} r_e {k^2} + O(k^4) - i k \right].
\end{equation}
Unitarity guarantees that the imaginary parts agree.  Equating the
real parts at $k=0$ yields

\begin{equation}
\frac{M}{4 \pi a}=\frac{(C_2 I_3 -1)^2}{C + C_2^2 I_5} - {I_1}.
\end{equation}

Equating the $k^2$ coefficients gives

\begin{equation}
\frac{M r_e}{8 \pi}=\left(\frac{M}{4 \pi a} + I_1\right)^2
\left[\frac{1}{(C_2 I_3 - 1)^2 I_3} - \frac{1}{I_3}\right].
\label{eq:recondn}
\end{equation}
Note that the physical parameters are complicated non-linear
functions of divergent integrals and bare parameters.  The scattering
amplitude can now be written as

\begin{equation}
\mbox{Re}\left(\frac{1}{T^{\rm on}(k)}\right)=\frac{M/(4 \pi a) - {k^2} I_1 A}
{1 + {k^2} A},
\label{eq:Tsimple}
\end{equation}
with

\begin{equation}
A \equiv {\frac{M r_e}{8 \pi}}(\frac{M}{4 \pi a} + I_1)^{-1}.
\label{eq:Adef}
\end{equation}
Notice that the scattering amplitude is now expressed in terms of the
physical parameters $a$ and $r_e$ and the linearly divergent integral
$I_1$. It is not surprising that we have a divergence left over after
imposing our renormalization conditions since we had three distinct
divergences---$I_1$, $I_3$ and $I_5$---and only two free
parameters---$C$ and $C_2$.

\subsection{Cutoff Regularization}
\label{sec-V2cutoff}

If we now regularize $I_1$ using a sharp cutoff,
$\beta$, we find in the $\beta\rightarrow\infty$ limit:

\begin{equation}
\frac{1}{T^{\rm on}(k)}=-\frac{M}{4 \pi}\left[ -\frac{1}{a} +
\frac{1}{2} r_e {k^2} - i k \right].
\label{eq:cutoff2}
\end{equation}
which is the effective-range expansion to order $k^2$. Given the
success of Eq.~(\ref{eq:cutoff2}) in describing the ${}^1S_0$ $NN$
phase shifts this result seems quite promising. However, there is a
subtlety hidden in this result. The second term in
Eq.~(\ref{eq:recondn}) disappears when the cutoff is removed.
Consequently, as $\beta \rightarrow \infty$

\begin{equation}
\frac{M r_e}{8 \pi} \rightarrow \frac{1}{I_3} \left(\frac{I_1}{C_2 I_3
- 1}\right)^2,
\end{equation}
from which we see that $r_e \leq 0$, (recalling that $I_n$ defined in
Eq.~(\ref{In}) is negative definite), {\it no matter what real value
we choose for the constant $C_2$ in the bare Lagrangian}. This result
is at first sight rather peculiar since it implies that the sign of
$r_e$ is predicted in the effective theory with the cutoff removed.
It is the behavior of the power-law divergences arising from integrals
over internal loop momenta which leads to the result $r_e \leq
0$. This is in contrast to the ``zeroth-order'' amplitude, where these
integrals over internal loop momenta did not affect the final on-shell
amplitude.  Moreover, the result $r_e \leq 0$ as $\beta \rightarrow
\infty$ clearly means that in order to describe a system with
${r_e}>0$ using this potential one cannot remove the cutoff but must
instead maintain the hierarchy $\beta\leq O(1/{r_e})$. (The necessity
of keeping the cutoff small and below $O(1/r_e)$ is supported by
direct calculation~\cite{Sc97}.)

\subsection{Wigner's Bound}
\label{sec-Wb}

The fact that one cannot take the cutoff $\beta$ to infinity and still
obtain a positive effective range is, in fact, not surprising.  Long
ago, assuming only causality and unitarity, Wigner proved that if a
potential vanishes beyond range $R$ then the rate at which phase
shifts can change with energy is bounded by~\cite{Wi55}

\begin{equation}
\frac{d \delta(k)}{d k} \, \geq \, - \, R \, + \, \frac{1}{2
k}\sin(2\delta(k) + 2 k R).
\end{equation}
It is straightforward to show that this translates to~\cite{PC97,Fe95}

\begin{equation}
{r_e}\leq 2\left[ R-\frac{R^2}{a}+\frac{R^3}{3{a^2}}\right].
\label{eq:wb}
\end{equation}
In light of this general result it is no longer surprising that the
effective range was found to be negative in the EFT calculation when
the cutoff $\beta$ was removed. Wigner's bound implies that no theory
obeying the general physical principles of causality and unitarity can
support a positive effective range with a zero-range force.  When
using a sharp cutoff the regulated Hamiltonian satisfies all physical
principles and is therefore a legitimate physical theory, with the
cutoff identified with the range; i.e. $\beta \sim 1/R$. Consequently
in the limit $\beta \rightarrow \infty$ one is trying to use
zero-range potentials to produce a positive effective range. Equation
(\ref{eq:wb}) shows that this is impossible. Therefore, Wigner's bound
indicates that there is no EFT treatment of $NN$ scattering with
nucleons as the only explicit degrees of freedom in which it is
possible to remove the cutoff from the problem. One might think that
including explicit pions in the effective theory description will
ameliorate the situation. This is not so; it has been shown that the
Wigner bound is a general feature of contact interactions and
continues to apply in the presence of pions~\cite{Sc97}.

\section{Dimensional Regularization}

As we have seen, {\it a priori} there are two sources of
short-distance physics in the effective field theory calculation:
operators in the effective Lagrangian of increasing dimensionality
whose coefficients encode information about short-distance physics,
and the effects of virtual particle excitations within loop graphs. In
ordinary perturbative EFT calculations the existence of a consistent
power-counting scheme in the S-matrix relies on removing the
short-distance physics that arises in loop graphs via the
renormalization procedure~\cite{We79}. Since the short-distance
physics is removed in a manner insensitive to choice of regularization
scheme it is economical to use DR to regularize and renormalize,
because DR respects chiral and gauge symmetries. When considering the
relevance of EFT methods in nuclear physics one might choose to
extrapolate intuition gained from perturbation theory and regulate
using DR. However, it is important to realize that the use of DR
implicitly {\it assumes} that EFT can work only if the short-distance
physics buried in loop graphs {\it does not} contribute to low-energy
physics. Of course there is no need to make this assumption, cutoff
regularization is the most physically-transparent method of
regularization and keeps track of all sources of short-distance
physics in the calculation.

In this section we compare DR and cutoff schemes.  There are two
fundamental points we wish to make: (i) DR and cutoff regularization
do not lead to the same physical scattering amplitude.  Therefore in
the nonperturbative context of $NN$ scattering, short-distance physics
from loops is important, in contrast to the situation in perturbative
calculations; (ii) When low-lying (quasi-)bound states are present, a
necessary (but not sufficient) condition for a workable EFT treatment
of $NN$ scattering is that short-distance effects from loops
contribute to the physical scattering amplitude.

We now use dimensional regularization (DR) to regulate the integrals
$I_1$, $I_3$, and $I_5$ which appear in the expression for
$T^{\rm{on}}$ above.  As discussed above, DR is a convenient way to
implement the idea that is central to the success of perturbative EFT:
the power-law divergent pieces of integrals over internal loop momenta
should not affect the final physical scattering amplitude. If this
holds true then DR should lead to the amplitude found using cutoff
regularization.  In DR all power-law divergences vanish, therefore
$I_1=I_3=I_5=0$.  Consequently, Eqs.~(\ref{eq:Tsimple}) and
(\ref{eq:Adef}) reduce to

\begin{equation}
\frac{1}{T^{\rm on}_{\rm DR}(k)}=
-\frac{M}{4\pi}\left[ \frac{1}{(-a - \frac{1}{2} r_e a^2 k^2 )}- ik \right],
\label{eq:TDR}
\end{equation}
in agreement with Kaplan {\it et al.}~\cite{Ka96}. The renormalized
scattering amplitude is not the same as that obtained using cutoff
regularization. Therefore the amplitude calculated using the
second-order EFT potential is not independent of regularization
scheme, and so the scattering amplitude must be sensitive to the
short-distance physics from loops.  This raises the question of
whether the EFT treatment based on potentials has really managed to
separate the long-distance physics from the short-distance
physics. This separation is a precondition for the successful
application of EFT ideas.

Clearly the DR amplitude maps to the effective-range expansion only
for momenta $k \ll 1/\sqrt{a{r_e}}$. This is consistent with
Eq.~(\ref{eq:cutoff2}) if both $a$ and $r_e$ are natural, as then
Eqs.~(\ref{eq:TDR}) and (\ref{eq:cutoff2}) will agree within the
domain of validity of the EFT, $k^2 \ll \Lambda^2$. The effective
theory is then perturbative.  However, if the scattering length is
unnaturally large the momentum domain within which the two forms are
equivalent becomes small. In fact, Eq.~(\ref{eq:TDR}) reproduces the
data in the ${}^1S_0$ extremely poorly, since it only agrees with the
phenomenologically efficacious amplitude (\ref{eq:reamp}) for very
small $k$.  It was argued elsewhere that this failure implies that the
effective theory calculation is not valid in the presence of a large
scattering length~\cite{Ka96}.

\begin{figure}[h,t,b]
   \vspace{0.5cm} 
   \epsfysize=2cm 
   \centerline{\epsffile{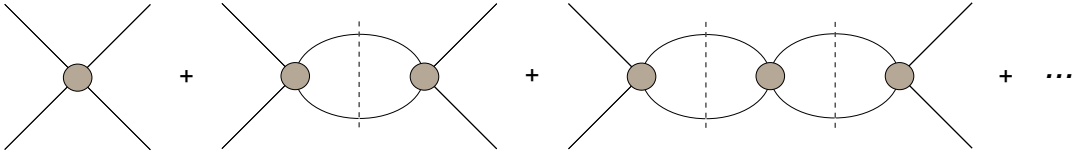}}
   \centerline{\parbox{11cm}
   {\caption{\label{fig3} The diagrammatic solution of the
dimensionally regulated LS equation with potential represented by the
shaded blob. Only the absorptive parts of the loop graphs remain.
}}}
\end{figure}

However, this failure of dimensional regularization is no surprise
when one considers the simple model of the effective-range expansion
discussed in Sec.~\ref{sec-sm}. There it was clear that an unnaturally
large scattering length occurs through the cancellation between two
terms in the scattering amplitude, one of which arises from
integration over loop momenta. This cancellation between ``range'' and
``strength'' is a general feature whenever one solves the
Schr\"odinger equation with some finite-range potential.  By its very
nature, DR {\it cannot} give a scattering amplitude that feels this
cancellation, since it discards all short-distance physics that comes
from loops.  In fact, since there are no logarithmic divergences in
this problem and the loop graphs have no finite real part, it is
straightforward to show that DR gives an amplitude in which only the
absorptive parts of the loop graphs are retained~\footnote{This point
has also recently been made by Lepage~\cite{Le97}.}. (See
Fig.~\ref{fig3}.)  Hence when there are low-lying bound states the
bound-state energy sets the scale of the coefficients in the potential
of the dimensional regularization calculation. But recall that this
was precisely the problem that iterating the potential via the LS
equation was supposed to avoid.  Thus the implementation of the
general cancellation between range and strength must be an element of
any EFT description of $NN$ scattering.

In general, information about the range of the potential enters
through these power-law divergences, and DR's neglect of all
power-law divergences means it discards this information on the range
of the interaction. This is no accident, DR is designed to be a
scale-independent regularization scheme. This aspect of the DR
prescription allows it to violate the Wigner bound and give an
amplitude which has a positive effective range from an EFT potential
which is apparently zero range.  

Therefore we argue that a regulator that introduces a new scale in
$NN$ scattering must be used. However, in the next section we will
show that effects due to this new scale actually destroy power
counting. Therefore, we believe that the intuition that EFT can work
only if scattering amplitudes are insensitive to the short-distance
physics arising from loop graphs, which is the primary motivation for
using DR, is actually correct. In other words, in all circumstances in
which EFT is systematic DR can, and for simplicity should, be used. In
the calculation of $NN$ scattering in the presence of low-energy bound
states DR fails only because the sensitivity to short-distance physics
inherent in this problem means that no sensible EFT can be formulated
for this problem.

\section{Power Counting in Cutoff Effective Field Theory}

\label{sec-coeft}

The arguments of the previous section show that an EFT approach to
$NN$ scattering will only be efficacious in the presence of low-lying
(quasi-)bound states if the short-distance physics in loop graphs
contributes to the physical scattering amplitude.  If a cutoff is
introduced, so that we can keep track of this physics, the bound
(\ref{eq:wb}) indicates that it cannot be taken to infinity.  Therefore,
in this section we investigate the possibility of an EFT for $NN$
scattering in the presence of a finite cutoff. 

This physically intuitive approach has been advocated by
Lepage~\cite{Le97,Le90}.  The idea is that one takes an underlying
theory of $NN$ interactions and introduces a (sharp or smooth)
momentum cutoff $\beta$ representing the scale at which the first new
physics becomes important. All loops now only
include momenta $p < \beta$. Of course, one must compensate for the
effects of these neglected modes. However, Lepage argues that since
these modes are highly virtual, one may approximate their effects by a
sequence of local contact interactions. Furthermore, if the cutoff
$\beta$ is placed well below the mass $\Lambda$ of some exchanged
quantum, then, for momenta ${\bf p}$ and ${\bf p}'$ below the cutoff,
the exchange of this quantum:
\begin{equation}
V_\Lambda({\bf p}',{\bf p}) \sim \frac{1}{({\bf p}' - {\bf p})^2 + \Lambda^2};
\end{equation}
may be replaced by a contact interaction, since $p',p < \beta \ll
\Lambda$.  Therefore the effects coming from exchanges of quanta with
masses well above the cutoff scale $\beta$ may also be approximated by
contact interactions. For the numerical application of these ideas to
the $NN$ problem see~\cite{Or96,Sc97,Le97}.

Now, all that has been said in the previous paragraph still applies if
we set our cutoff $\beta$ below the scale $m_\pi$. Then the only
explicit degrees of freedom in the problem are nucleon modes with
momentum $\beta < m_\pi$. All higher-momentum nucleon modes {\em and}
all exchanged mesons are integrated out. This cutoff effective field
theory of the $NN$ interaction is of little practical use, but can be
investigated analytically in a way that raises issues of principle. The
effective Lagrangian is

\begin{equation}
{\cal L}=N^\dagger i \partial_t N - N^\dagger \frac{\nabla^2}{2 M} N
- \frac{1}{2} C(\beta)  (N^\dagger N)^2\\ 
-\frac{1}{2} C_2(\beta) (N^\dagger \nabla^2 N) (N^\dagger N) 
+ h.c.+\ldots
\label{eq:colag}
\end{equation}
where the dots signify operators with more than two derivatives. As
discussed in Sec.~\ref{sec-eft}, in the limit that the nucleon
is very heavy we can formally calculate the four-point function
exactly.  To do this we merely write down the effective potential
which corresponds to this Lagrangian:

\begin{equation}
V(p',p)=[C + C_2 (p^2 + p'^2) + \ldots] \theta(\beta - p)
\theta(\beta - p'),
\label{eq:Veffexp}
\end{equation}
and iterate via the Lippmann-Schwinger equation (\ref{eq:LSE2}).  Here
the theta function imposes a sharp cutoff, and so all integrals
(not just the divergent ones) will be cut off sharply at momentum
$\beta$. After renormalization the coefficients $C$, $C_2$, etc., will, of 
course, depend on the cutoff scale $\beta$, as well as physical
scales in the problem. 

Of course, the expression (\ref{eq:Veffexp}) is an infinite series,
and for practical computation some method of truncating it must be
found.  The fundamental philosophy of cutoff EFT provides a rationale
for this as follows. If we work to any finite order in the effective
potential, cutoff-dependent terms in the scattering amplitude will
appear. These must be in correspondence with neglected higher-order
operators in $V$.  If such terms are progressively added to
$V$, one may remove the cutoff dependence
order-by-order. Below we will see that one can indeed define such a
``systematic'' procedure in this problem. However, {\it in order that
it really make sense to truncate the effective potential
(\ref{eq:Veffexp}) at some finite order it must be that the operators
that are neglected are in some sense small.}

Naively one would expect that a sufficient condition for the operator with
$n$ derivatives to be a small correction to the overall potential is:

\begin{equation}
\hat{p}^{2(n-m)} \ll \frac{C_{2m}}{C_{2n}}; \quad \forall \; m \geq 0 \mbox{
such that } n > m,
\label{eq:pcondn}
\end{equation}
where $\hat{p}$ is the momentum operator. This operator yields some
characteristic momentum in the problem. Now, in perturbation theory
the short-distance physics can be removed from loops via
renormalization, thereby achieving a clean separation of scales, and
so the characteristic momentum is simply the external momentum. Thus,
in perturbative calculations the operator hierarchy (\ref{eq:pcondn})
can be maintained, as long as one restricts oneself to small enough
external momenta.

However, in a nonperturbative context we have already seen that
physics from loops must enter into the calculation if low-lying
(quasi-)bound states are to be generated. Consequently, we expect that
momenta at scales $\hat{p}^2\sim \beta^2$ are important. Below we show
that
\begin{equation}
\frac{C_2}{C} \sim \frac{1}{\beta^2}.
\label{eq:CC2}
\end{equation}
More generally, condition (\ref{eq:CC2}) becomes:
\begin{equation}
\frac{C_{2m}}{C_{2n}} \sim \beta^{2(n-m)}.
\label{eq:CmCnbeh}
\end{equation}
Combining Eqs.~(\ref{eq:CmCnbeh}) and (\ref{eq:pcondn}) we obtain
\begin{equation}
(\hat{p}/\beta)^j \ll 1; \quad j=2,4,6,\ldots.
\label{eq:success}
\end{equation}
Equation (\ref{eq:success}) reflects the fact that the crucial momenta
in the problem must be well below the scale where the cutoff effective
field theory breaks down. The external momentum can certainly be kept
in this energy regime. However, below we also show that, even if the
external momentum scales are small, the quantum mechanical averages of
operators $\langle \hat{p}^{j} \rangle$ involve $\beta$. In this
circumstance there is little justification for truncating the
effective Lagrangian and keeping only a finite number of operators.

\subsection{``Leading order'' cutoff EFT calculation ($\nu=0$)}
\label{sec-nu0ceft}

Consider once again the ${}^1S_0$ $NN$ amplitude generated by the
``zeroth-order'' EFT potential.  This potential in cutoff EFT is given
by

\begin{equation}
V^{(0)}(p',p)=C^{(0)}\theta(\beta - p) \theta(\beta-p')
\label{eq:V0coeft}
\end{equation}
The solution of the Lippmann-Schwinger equation is then

\begin{equation}
\frac{1}{T^{\rm on}(k)}=\frac{1}{C^{(0)}} - {I_\beta}(k),
\end{equation}
where 

\begin{equation}
I_\beta(k)=-M \int \frac{d^3q}{(2 \pi)^3} \frac{1}{k^2 - q^2+i\epsilon}
\theta(\beta - q).
\end{equation}
Matching the solution to the effective-range expansion gives

\begin{equation}
\frac{1}{T^{\rm on}(k)}=-\frac{M}{4 \pi}\left[ -\frac{1}{a} +
\frac{1}{2} r^{(0)}_e {k^2} + O(k^4) - i k \right],
\end{equation}
where the renormalization condition for $C^{(0)}$ is still
Eq.~(\ref{eq:T0cutoff}). Note that $a$ is cutoff independent as
before, but now we have a non-zero effective range in this
zeroth-order calculation:

\begin{equation}
r^{(0)}_e=\frac{4}{\pi\beta}.
\end{equation}
This $r_e^{(0)}$ arises from the energy dependence introduced by the
cutoff. The appearance of a cutoff-dependent quantity like $r_e^{(0)}$
in the renormalized amplitude only reflects the limited accuracy of
this calculation. Below we will see that $r_e$ can be made
cutoff independent by including higher-order terms in the effective
potential.

\subsection{``Next-to-leading Order'' cutoff EFT calculation($\nu$=2)}
\label{sec-nu2ceft}

We now turn our attention to the amplitude and renormalization
conditions that arise when one takes the ``second-order'' effective
potential:
\begin{equation}
V^{(2)}(p',p)=[C + C_2 (p^2 + p'^2)] \theta(\beta - p)
\theta(\beta - p'),
\end{equation}
and iterates it via the Lippmann-Schwinger equation.
Repeating the analysis of Sec.~\ref{sec-nu2}, but now with all
integrals cutoff sharply at momentum $\beta$, we again obtain the amplitude:
\begin{equation}
\frac{1}{T^{\rm on}(k)}=\frac{(C_2 I_3 -1)^2}{C + C_2^2 I_5 + {k^2} C_2 (2 -
C_2 I_3)} - I_\beta(k),
\end{equation}
while $I_3$, and $I_5$ are as before, provided that it is understood
that a sharp cutoff is to be used to render the divergences finite.
>From these equations we obtain the following equations for $C$ and
$C_2$:
\begin{eqnarray}
\frac{M}{4 \pi a}&=&\frac{(C_2 I_3 -1)^2}{C + C_2^2 I_5} - {I_1};
\label{eq:Cbeta}\\
\frac{M r_e}{8 \pi}&=&\left(\frac{M}{4 \pi a} + I_1\right)^2
\frac{C_2 (2 - C_2 I_3)}{(C_2 I_3 - 1)^2} + \frac{M r_e^{(0)}}{8 \pi},
\label{eq:C2beta}
\end{eqnarray}
where, as above, the last term in Eq.~(\ref{eq:C2beta}) arises because
the presence of the cutoff generates additional energy dependence when
the integral $I_\beta(k)$ is evaluated. 

Of course, as $\beta$ is varied the $C$ and $C_2$ that satisfy
Eqs.~(\ref{eq:Cbeta}) and (\ref{eq:C2beta}) will change
significantly. However, because one is fitting to low-energy
scattering data different values of $\beta$ will not lead to any
fundamental differences in the low-energy T-matrix. Since $C$ and
$C_2$ are fit to the first two terms in the effective-range expansion
sensitivity to the cutoff enters first through the fourth-order term
in the effective-range expansion.  Therefore the effects of choosing
different cutoffs can only appear in the on-shell amplitude at order
$(k/\beta)^4$, where $k$ is the on-shell momentum.

This is precisely the sense in which working to higher order in the
potential will improve the fit ``systematically''.  If we constructed
the ``next-order'' effective potential, $V^{(4)}$ and refitted the
(four) coefficients appearing in it to the low-energy data then
sensitivity to the cutoff will be pushed back to an $O((k/\beta)^8)$
correction. Thus, the introduction of higher-dimension operators
allows for the ``systematic'' removal of cutoff dependence in the
amplitude. In general, the two-body scattering amplitude will be
insensitive to the choice of regulator, provided one restricts oneself
to the domain of validity of the cutoff effective field theory, $k \ll
\beta$.

This implies that so long as we renormalize the coefficients to the
low-energy scattering data we ``systematically'' improve our fits to
the scattering data. This is the systematicity seen in the work of
Lepage~\cite{Le97}, and that of Scaldeferri {\it et
al.}~\cite{Sc97}. Nevertheless, this does not yield a systematic EFT
potential in the sense we have defined it here.  After all, {\it any}
parameterization of a potential which is rich enough to fit the lowest
terms in the effective-range expansion will similarly be improved as
one adds parameters to it. The question is whether one can use the
power counting to argue {\it a priori} that certain contributions to
the potential of Eq.~(\ref{eq:Veffexp}) will be systematically small and
hence can be neglected at some specified level of accuracy.

To answer this question we must examine the values of $C$ and $C_2$
which are required to solve the equations above.  Assuming that we are
working in the regime where $1/a \ll \beta$ it follows that
the second of these two equations becomes
\begin{equation}
\frac{M}{8 \pi}(r_e - r_e^{(0)}) \approx \frac{I_1^2}{I_3}
\frac{\overline{C}_2 (2 - \overline{C}_2)}{(\overline{C}_2 - 1)^2},
\end{equation}
where $\overline{C}_2 \equiv C_2 I_3$. This leads to a quadratic
equation for $\overline{C}_2$, which for values of $\beta$ up to some
$\beta_{\rm max}$ has real solutions. For $\beta > \beta_{\rm max}$
the renormalization condition for $C_2$ has no real solution if $r_e
>0$, as discussed in Sec.~\ref{sec-nu2} and Ref.~\cite{Ph97}. However,
if $\beta \ll 4/(\pi r_e)$, then we see that $\overline{C}_2=1 \pm
\frac{\sqrt{3}}{2}$.  Consequently, we infer
\begin{equation}
C_2 \sim \frac{1}{M \beta^3} \Rightarrow C \sim \frac{1}{M \beta}.
\label{eq:CC2beh}
\end{equation}

This behavior for $C$ is the same as that obtained in the zeroth-order
calculation in the case $1/a \ll \beta$. Since we demand that the
physical observable $a$ be the same in both the zeroth and
second-order calculations, the renormalization conditions
(\ref{eq:T0cutoff}) for $C^{(0)}$ and (\ref{eq:Cbeta}) for $C$ give
\begin{equation}
\overline{C}^{(0)}=\frac{1}{(1-\overline{C}_2)^2}
\left(\overline{C} + \frac{9}{5} \overline{C}_2^2 \right),
\label{eq:CC0comp}
\end{equation}
where $\overline{C}$ ($\overline{C}^{(0)}$) is $I_1$ times the
original dimensionful coefficient. Now $\overline{C}_2 \equiv C_2I_3=1
\pm \frac{\sqrt{3}}{2}$, provided that $\beta \ll 4/(\pi r_e)$. Thus,
Eq.~(\ref{eq:CC0comp}) shows that if $1/a \ll \beta \ll 4/(\pi r_e)$,
$\overline{C}$ differs significantly from the lowest-order
result. Hence, if both the zeroth and second-order effective
potentials are to reproduce the scattering length then there must be a
large difference between them.  This already suggests that a
systematic truncation of the cutoff EFT may not be possible.

Note that such problems do not arise if $a$ is natural, i.e. of order
$1/m_\pi$, and $\beta$ is chosen to be much less than $m_\pi$; then
the leading order behavior of the coefficients $C$ and $C_2$ is very
different. In fact,
\begin{equation}
C \sim \frac{1}{M m_\pi}; \qquad C_2 \sim \frac{1}{M m_\pi^2 \beta}.
\end{equation}
In this case all loop effects coming from $C_2$ are suppressed by a
factor of at least $(\beta/m_\pi)^2$. Indeed, all loop effects are
suppressed by a factor of at least $\beta/m_\pi$.  Therefore, if
$\beta \ll m_\pi$ a non-perturbative calculation is not necessary. In
other words, if the experimental parameters are natural then cutoff
field theory with $\beta \ll m_\pi$ gives a perturbative EFT in which
loop graphs are consistently suppressed~\cite{Ma95}. However, if a
perturbative calculation is performed then the regularization scheme
chosen becomes immaterial, as the short-distance physics may be
renormalized away. Dimensional regularization would be an equally
valid, and considerably simpler, way of implementing a systematic
perturbative EFT. Therefore we do not consider the case where $1/a$ is
of order $m_\pi$ any further here, but instead return to the case
where $1/a$ is unnaturally small, as is true in the ${}^1S_0$ and
${}^3S_1-{}^3D_1$ channels in nature.

For unnaturally long scattering lengths Eq.~(\ref{eq:CC2beh}) shows that:
\begin{equation}
\frac{C_2}{C} \sim \frac{1}{\beta^2}.
\label{eq:ratio}
\end{equation}
Consequently the condition (\ref{eq:pcondn}) becomes, for $n=2$, $\hat{p}^2
\ll \beta^2$.  It is easy to give a heuristic justification of why
this behavior of the coefficients arises in a non-perturbative cutoff
EFT calculation, and why we expect similar behavior to all orders in
the effective potential. After all, the choice of a theta function to
regulate the momentum-space integrals as in Eq.~(\ref{eq:Veffexp}) is
entirely arbitrary. All that has been said above could be reformulated
with a smooth cutoff. This would result in an effective potential of
the form
\begin{equation}
V(p',p)=[\tilde{C} + \tilde{C_2} (p^2 + p'^2) + \ldots]
g(p^2/\beta^2,p'^2/\beta^2)
\label{eq:Veffexp2}
\end{equation}
where $g(x,y)$ obeys $g(0,0)=1$, $g(x,y)=g(y,x)$ and $g(x,y)
\rightarrow 0$ faster than any power of $x$ as $x \rightarrow \infty$
with $y$ held fixed. In a non-perturbative calculation the effective
potential should be essentially unaltered by this change in the form
of the cutoff. However, this necessarily means that the ratios
$\tilde{C_{2n}}/\tilde{C}$ differ from those $C_{2n}/C$ by terms of order
$1/\beta^{2n}$.  Therefore for a generic cutoff function $g$ the 
ratio $C_{2n}/C$ must be of order $1/\beta^{2n}$.

Now, if the ratio $C_{2n}/C$ goes like $1/\beta^{2n}$, then the
condition (\ref{eq:pcondn}) becomes Eq.~(\ref{eq:success}). However,
the effective potential is to be used in a momentum regime which
extends up to $\beta$, and at the upper end of this momentum regime it
is clear that all terms in the expansion for $V$ are
equally important.

Of course, if internal loops were dominated by the external momentum,
$k$, and so $\hat{p} \approx k$, then this behavior of the
coefficients would not be cause for concern, since $k \ll \beta$ could
be maintained. However, virtual momenta up to $\beta$ flow through all
internal loops, and we have already argued that the final amplitude is
sensitive to these virtual effects which come from the range of the
underlying interaction. Therefore we believe it makes sense to
consider a quantum average in evaluating the condition
(\ref{eq:pcondn}), since such an average is sensitive to these virtual
effects.

Although there is no bound state in this channel, all arguments about
the size of operators in the effective action would apply equally well
if there was a low-energy bound state in the channel under
consideration. So, let us evaluate quantum averages of the operator
$\hat{p}^{2n}$ using the bound-state wave function obtained from the
zeroth-order potential given by Eq.~(\ref{eq:V0coeft}). We use a
bound-state wave function because scattering wave functions are not
easily normalizable, and so quantum averages become difficult to
calculate.  The zeroth-order potential yields a wave function for the
bound state of energy $E=-B$,
\begin{equation}
\psi^{(0)}(p)={\cal N} \frac{M}{MB + p^2} \theta(\beta-p),
\end{equation}
where ${\cal N}$ is some normalization constant that is yet to be
determined. In the regime $MB \ll \beta^2$ this wave function gives
\begin{equation}
\frac{\langle \hat{p}^{2n} \rangle}{\beta^{2n}} \equiv \frac{\langle
\psi^{(0)}|\hat{p}^{2n}|\psi^{(0)} \rangle}{\beta^{2n}} = \frac{4}{(2n-1) \pi}
\frac{\sqrt{MB}}{\beta}, \qquad n=1,2,\ldots.
\end{equation}
Thus, condition (\ref{eq:success}) is apparently satisfied. However,
if this condition is to hold as an operator equation it must be
that, for any $n>0$:
\begin{equation}
\frac{1}{\beta^{2(n-m)}} \frac{\langle
\psi^{(0)}|\hat{p}^{2n}|\psi^{(0)} \rangle}{\langle
\psi^{(0)}|\hat{p}^{2m}|\psi^{(0)} \rangle} \ll 1
\quad \forall \; m \geq 0 \mbox{ such that } n > m.
\end{equation}
For $m>0$ this does not hold. That is to say, if we calculate $\langle
V \rangle$ with the wave function $\psi^{(0)}$ there is no reason to
truncate the expansion at any finite order, since all terms beyond
zeroth order contribute with equal strength to the quantum average.

If there was systematic power counting for the $NN$ potential in
cutoff field theory then the contribution of these ``higher-order''
terms in the potential should get systematically smaller as the
``order'' is increased.  However, it is clear that this does not
happen---rather, all terms beyond zeroth order contribute to the
potential at the same order.  Therefore one cannot justify a
truncation of Eqs.~(\ref{eq:Veffexp}) and (\ref{eq:Veffexp2}) at some
finite order in $p$ and $p'$. For the reasons explained above, such a
truncation may result in a good fit to the experimental data for
on-shell momenta $k \ll \beta$, but it is not based on a systematic
expansion of the $NN$ potential in powers of momentum.

\section{Discussion}
\label{sec-discuss}

Our results are discouraging for any effective field theory
description of $NN$ scattering in which low-lying bound states are not
included as explicit degrees of freedom. These low-energy bound and
quasi-bound states exist in nuclear physics, and so, if EFTs in which
they are not explicitly included are to be useful, power-counting
arguments can only apply to the $NN$ potential, rather than to the
$NN$ amplitude. Solving such an EFT with nucleons would appear to be a
simple problem in non-relativistic quantum mechanics. However, the
problem is complicated by the appearance of power-law divergences
which arise from the internal loop integrations in the
Lippmann-Schwinger equation. In perturbative EFTs such divergences are
renormalized away and have no effect on the physical amplitude. In
fact, if this were not the case there would be no consistent power
counting in these theories.  By contrast, when the second-order EFT
potential is iterated to all orders, we find that the resulting
scattering amplitude is inevitably regularization scheme dependent:
dimensional regularization and cutoff regularization give different
renormalized scattering amplitudes.  We are therefore forced to
conclude that the physical scattering amplitude is sensitive to the
power-law divergences arising from loop integrations. One way to
understand this sensitivity is to realize that when cutoff
regularization is used the potential acquires a range, which is
naively $1/\beta$.  When the Schr\"odinger equation is solved all
physical observables depend on the upper limit in internal loop
integrations, $\beta$. In any perturbative calculation this dependence
can be removed by the introduction of a finite number of
higher-dimensional operators in the Lagrangian. However, for some
observables, e.g. the effective range, this cannot be achieved for
arbitrary cutoff in our non-perturbative calculation: these
observables still depend on the cutoff (equivalently, on the ``range''
of the potential), even after renormalization, as evidenced by the
bound (\ref{eq:wb}). When dimensional regularization discards all
power-law divergences in order to maintain its scale independence it
removes, by fiat, the power-law divergences which represent an
important part of the physics of the range of the $NN$
interaction. This suggests DR should not be used to regularize this
potential.

It is straightforward to show that if DR is used to regularize a
potential which generates low-energy bound or quasi-bound states, the
resulting coefficients in the dimensionally regularized potential will
be governed by the bound-state energy, and not by the underlying scale
of the EFT~\cite{LM97}. Thus, if DR is used to regularize the EFT $NN$
potential the radius of convergence of the resulting expansion will be
small---as seen in the work of Kaplan {\it et al.}~\cite{Ka96}. Again,
this may be thought of as a consequence of DR's discarding power-law
divergences: such power-law divergences implement the cancellation
between ``range'' and ``strength'' which was seen in the simple model
of Sec.~\ref{sec-sm} (see also Ref.~\cite{Ka97}) to be the key to
allowing a theory with natural scales to produce ``unnaturally''
low-lying bound states. {\it Therefore the only hope for a systematic
and successful EFT description of $NN$ scattering with nucleons alone
lies with methods of regularization which preserve information on
these power-law divergences.}  Cutoff regularization is such a
regularization scheme.  However, the effective range is positive in
nature, and so we discover that if we wish to use an effective field
theory with a cutoff to describe the real world we must keep the
cutoff finite and below the scale $\Lambda$ of the short-distance
physics. While this may provide a procedure for fitting low-energy
scattering data, such an approach can only be regarded as systematic
if the important momenta inside internal loops are well below the
cutoff scale. If this is not true physical observables will be
sensitive to details of the potential in precisely the region where
the artificially-introduced cutoff plays an important role. As we saw
in Sec.~\ref{sec-coeft} there will then be no justification for
keeping some terms in the EFT potential and neglecting others. Now,
while the external momentum $k$ can clearly be kept small compared to
the cutoff, we have just argued that in order to have a successful EFT
in the presence of low-lying bound states the virtual momenta of order
the cutoff must play a significant role! This becomes apparent
mathematically through the strong cutoff dependence of the quantum
averages of momentum operators. {\it We therefore conclude that there
is no small parameter which allows one to consistently truncate the
regularized $NN$ potential at some finite order in the derivative
expansion.}

The preceding argument that power counting fails might be regarded as
incomplete since we have demonstrated a violation of power counting at
the level of the bare parameters in the regularized potential. In
general one does not expect power counting in effective field theories
to be respected at the level of the bare parameters in the Lagrangian
but rather at the level of the renormalized parameters. For example,
in ordinary chiral perturbation theory, the bare parameters which
enter the perturbative calculations are strictly infinite, while the
renormalized amplitudes have natural power counting.  Here we have
shown that sensible regularization schemes inevitably involve a
finite cutoff and therefore the only object in which systematic power
counting can be defined is the potential with bare parameters; {\it
there simply is no renormalized potential}.  One might attempt to {\it
define} the renormalized potential as the $K$ matrix, $({\rm
Re}[T^{-1}])^{-1}$. This quantity is certainly well defined. However,
as discussed elsewhere in this paper, the entire reason for doing EFT
at the level of the potential is that we are interested in
developing an EFT in the vicinity of the low-energy pole, and power
counting in $T$ or $K$ fails in the presence of this pole.

It might also be suggested that the inclusion of pions as explicit
degrees of freedom in the effective field theory will ameliorate the
situation.  However, we do not believe this to be the case. After all,
provided the energy under consideration is low enough, nothing in the
general EFT arguments applied to the $NN$ potential in
Refs.~\cite{We90,We91,Ka96} requires the inclusion of explicit
pions. When pions are added as fields in the Lagrangian of the EFT the
basic object of the calculation remains an EFT potential which is to
be iterated via the Lippmann-Schwinger equation. This potential still
requires regularization. If cutoff regularization is used a
generalization of the Wigner bound implies that even in this EFT with
explicit pions the cutoff cannot be taken to
infinity~\cite{Sc97,PC97}. DR could also be used to render finite the
divergent integrals which arise upon iteration, but again, doing this
implicitly assumes that details of the short-distance physics do not
affect the physical scattering amplitude. (In fact, additional
problems, beyond those discussed here, arise when DR is applied to the
EFT with pions, as noted in Ref.~\cite{Ka96}.) Therefore it seems that
cutoff EFT, with a finite cutoff $\beta$, is the safest way to
regularize the EFT with explicit pions. Whether the inclusion of
explicit pions modifies our conclusions about cutoff EFT is a matter
for numerical investigation. Previous numerical calculations indicate
that the inclusion of explicit pions does not change the scales of the
coefficients $C$ and $C_2$~\cite{Sc97}. Hence, it does not appear that
the explicit inclusion of pions in the EFT will modify our conclusion
that there is no systematic EFT for $NN$ scattering.

Richardson {\it et al.}  have independently argued that the EFT
approach to $NN$ scattering is not systematic~\cite{Ri97}. However,
their definition of systematic is more restrictive than ours. They
define systematicity to mean that the scattering length $a$ should
receive no contribution from the term with two derivatives in the
effective Lagrangian (the coefficient $C_2$). Since $a$ does receive a
contribution from this term (see Eq.~(\ref{eq:Cbeta})) they claim that
the EFT (\ref{eq:lag}) is not systematic. This definition of
systematicity is patterned on perturbative EFT, and may be unduly
rigid.  We would argue that the appearance of $C_2$ in $a$ does not of
itself imply non-systematicity. After all, the contribution of $C_2$
to $a$ might be systematically small. This happens, for instance, if a
finite cutoff $\beta \ll m_\pi$ is used to regularize the divergences
and the physical observables are all natural. On the other hand, in
the case of interest here, where $1/a$ is unnaturally large, we saw in
the previous section that the contribution of the
appropriately-rescaled $C_2$ to $a$ {\it is} of order one. The
coefficients $C$ that are obtained in the zeroth and second-order
calculations then differ significantly.  Perhaps this is only
symptomatic of the more general malaise already diagnosed: the
higher-order terms in the effective action cannot be safely
ignored. In summary, the appearance of higher-order coefficients
in the expression for low-momentum observables does not necessarily
imply non-systematicity.  But here we expect the contribution of these
higher-order operators to low-momentum observables to be large,
because the effective potential cannot be systematically truncated.

The potential of EFT for the $NN$ interaction lies partly in embedding
the $NN$ scattering amplitude derived in the EFT in scattering
processes involving three or more nucleons. One would hope to apply
power-counting arguments to determine those contributions that are
important in the nuclear force and those that are systematically
suppressed. In principle, this approach has predictive power.  For
example, as noted by Weinberg, in a nucleus three-body forces are
characteristically down by two powers of $p/\Lambda$ compared to
two-body forces, and so might be expected to yield small contributions
to nuclear observables.  Furthermore, four-body forces are down by
four powers of $p/\Lambda$ and thus could be expected to be
negligible~\cite{We90,We91}. This suggests that at some reasonably
crude level one might simply neglect $N$-body force effects, $N \geq
3$ in nuclei. An analysis of realistic phenomenological nuclear force
calculations~\cite{Fr96} and the relative contributions of $N$-body
forces to nuclear binding energies shows qualitative agreement with
this prediction. Unfortunately, in light of the results of this paper
one must question in what sense it is valid to use power counting to
deduce a hierarchy of $N$-body forces. If EFT power counting is a
valid way to deduce that certain terms in the Hamiltonian are
suppressed then we would have seen that higher-derivative terms in the
potential were negligible. In fact, in none of the regularization
schemes discussed here was this the case.

Nevertheless, we stress that effective field theory continues to be
the most promising method of systematizing nuclear physics.  What we
have found here is that an EFT for $NN$ scattering with only nucleons
must contain {\it all} operators in the effective action, and
therefore is not useful. This has a simple physical
interpretation. The $NN$ scattering amplitude is sensitive to 
all operators in the effective action because it has a singularity
corresponding to the (quasi-)bound state pole. By definition, this
pole ``feels'' all distance scales and therefore naturally requires
that operators to all orders in momentum in the effective action be
present in order to describe it. The only way around this dilemma is
to include the (quasi-)bound state pole as a ``fundamental'' degree of
freedom in the effective theory, while retaining the four-point $NN$
interactions of the Lagrangian (\ref{eq:lag})~\cite{Ka97}.  In the
resulting EFT, these $NN$ contact interactions can be treated
perturbatively with consistent power counting. Application of these
ideas in the three-nucleon scattering problem is being pursued in
Ref.~\cite{Bd97}.

\section*{Acknowledgments}
We thank M.~K.~Banerjee, M.~C.~Birse, W.~D.~Linch, M.~A.~Luty, J.~A.~McGovern,
E.~F.~Redish, U.~van Kolck, S.~J.~Wallace, and S.~Weinberg for useful
and enjoyable discussions on the topics discussed in this paper. This
research was supported in part by the U. S. Department of Energy,
Nuclear Physics Division (grant DE-FG02-93ER-40762).


\end{document}